\documentclass[sigconf]{acmart}

\usepackage{booktabs} 
\usepackage{subfigure}
\usepackage{balance}
\usepackage{multirow}

\AtBeginDocument{%
  \providecommand\BibTeX{{%
    \normalfont B\kern-0.5em{\scshape i\kern-0.25em b}\kern-0.8em\TeX}}}

\copyrightyear{2020}
\acmYear{2020} 
\setcopyright{iw3c2w3}
\acmConference[WWW '20]{Proceedings of The Web Conference 2020}{April 20--24, 2020}{Taipei, Taiwan} 
\acmBooktitle{Proceedings of The Web Conference 2020 (WWW '20), April 20--24, 2020, Taipei, Taiwan}
\acmPrice{}
\acmDOI{10.1145/3366423.3380163}
\acmISBN{978-1-4503-7023-3/20/04}

\begin{document}

\title{Adversarial Multimodal Representation Learning for Click-Through Rate Prediction}

\author{Xiang Li, Chao Wang, Jiwei Tan, Xiaoyi Zeng, Dan Ou, Bo Zheng}
\affiliation{
  \institution{Alibaba Group, Hangzhou \& Beijing, China}
}
\email{{leo.lx, xiaoxuan.wc, jiwei.tjw, yuanhan, oudan.od, bozheng}@alibaba-inc.com}


\begin{abstract}
For better user experience and business effectiveness, Click-Through Rate (CTR) prediction has been one of the most important tasks in E-commerce. Although extensive CTR prediction models have been proposed, learning good representation of items from multimodal features is still less investigated, considering an item in E-commerce usually contains multiple heterogeneous modalities. Previous works either concatenate the multiple modality features, that is equivalent to giving a fixed importance weight to each modality; or learn dynamic weights of different modalities for different items through technique like attention mechanism. However, a problem is that there usually exists common redundant information across multiple modalities. The dynamic weights of different modalities computed by using the redundant information may not correctly reflect the different importance of each modality. To address this, we explore the complementarity and redundancy of modalities by considering modality-specific and modality-invariant features differently. We propose a novel Multimodal Adversarial Representation Network (MARN) for the CTR prediction task. A multimodal attention network first calculates the weights of multiple modalities for each item according to its modality-specific features. Then a multimodal adversarial network learns modality-invariant representations where a double-discriminators strategy is introduced. Finally, we achieve the multimodal item representations by combining both modality-specific and modality-invariant representations. We conduct extensive experiments on both public and industrial datasets, and the proposed method consistently achieves remarkable improvements to the state-of-the-art methods. Moreover, the approach has been deployed in an operational E-commerce system and online A/B testing further demonstrates the effectiveness.
\end{abstract}




\maketitle
\section{Introduction}

Large E-commerce portals such as Taobao and Amazon are serving hundreds of millions of users with billions of items. For better user experience and business effectiveness, Click-Through Rate (CTR) prediction has been one of the most important tasks in E-commerce. As widely studied by both academia and industry, extensive CTR prediction models have been proposed. Nevertheless, it is also an effective way to improve the CTR prediction accuracy through better mining and leveraging the multimodal features of items, since an item in E-commerce usually contains multiple heterogeneous modalities including IDs, image, title and statistic. Therefore, in this paper we make our effort to improve the CTR prediction accuracy by learning better representations for items of multiple modalities.

To leverage the multiple modalities for better item representations, a straightforward way~\cite{zhou2018deep,ni2018perceive,zhou2019deep} is to concatenate the multiple modality features, which is equivalent to giving a fixed importance weight to each modality regardless of different items. However, the importance of a modality may be different according to different items, and ideal item representations should be able to weigh different modalities dynamically so that emphasis on more useful signals is possible. For example, in the clothing category, whether users will click an item or not is highly affected by observing the images, so greater importance should be given to the image feature. On the contrary, in the cell phone and grocery food categories, the statistic feature of items reflects the popularity of items, while there is little difference between the images of items. A conceivable improvement~\cite{li2018multi,wu2019neural} is to learn dynamic weights of different modalities and emphasis on more useful signals through technique like attention mechanism. However, a problem is that there usually exists common redundant information across multiple modalities. The dynamic weights of different modalities computed by using the redundant information may not correctly reflect the different importance of each modality. 

Due to the above reason, our motivation is that the weight of completely independent information across modalities should be dynamic, and the weight of information shared by different modalities should be fixed. To address this, we divide multiple modality features into modality-specific (exist in one modality, and should have dynamic weights) and modality-invariant features (redundant in different modalities, and should have fixed weights). Take a dress which is displayed by an image with a title \emph{Girls Ballet Tutu Zebra Hot Pink} for example. The item involves a latent semantic feature of the material, such as yarn, which can be expressed by its image while not involved in its title, so the material feature is considered as the modality-specific (latent) feature for this example. The item also involves a common latent semantic feature of color (hot pink) in the subspace of both its image and title features, so the color feature is considered as the modality-invariant (latent) feature for this example. The key idea is that modality-specific features provide an effective way to explore dynamic contributions of different modalities, while modality-invariant features should have a fixed contribution and can be used as supplementary knowledge for comprehensive item representations. 

To the best of our knowledge, this is the first work that learns multimodal item representations by exploiting modality-specific and modality-invariant features differently. To achieve this, we propose a Multimodal Adversarial Representation Network (MARN) to deal with this challenging task. In MARN, a modality embedding layer extracts embedding vectors from multiple modalities and decomposes each embedding vector into modality-specific and modality-invariant features. Then, a multimodal attention network calculates the weights of multiple modalities for each item according to its modality-specific features. Also, a multimodal adversarial network learns modality-invariant representations where a double-discriminators strategy is introduced. The double-discriminators strategy is designed to identify the potential modalities involving common features across modalities and drive knowledge transfer between modalities. Finally, we achieve multimodal item representations by combining both modality-specific and modality-invariant representations. The contributions of the paper are summarized as:

\begin{itemize}
\item{The proposed MARN introduces a novel multimodal representation learning method for multimodal items, which can improve the CTR prediction accuracy in E-commerce.}

\item{We explore the complementarity and redundancy of modalities by considering modality-specific and modality-invariant features differently. To achieve discriminative representations, we propose a multimodal attention fusion network. Moreover, to achieve common representations across modalities, we propose a double-discriminators multimodal adversarial network.}

\item{We perform extensive experiments on both public and industrial datasets. MARN significantly outperforms the state-of-the-art methods. Moreover, the approach has been deployed in an operational E-commerce system and online A/B testing further demonstrates the effectiveness.}

\end{itemize}

\section{Related Work}

\subsection{Multimodal Learning}
Representing raw data in a format that a computational model can work with has always been a big challenge in machine learning~\cite{baltruvsaitis2018multimodal}. Multimodal representation learning methods aim to represent data using information from multiple modalities. Neural networks have become a very popular method for unimodal representations~\cite{bengio2013representation,elkahky2015multi}. They can represent visual or textual data and are increasingly used in the multimodal domain~\cite{ngiam2011multimodal,ouyang2014multi}. To construct multimodal representations using neural networks, each modality starts with several individual neural layers followed by a hidden layer that projects the modalities into a joint space~\cite{wu2014exploring}. The joint multimodal representations are then passed through extra multiple hidden layers or used directly for prediction.

The above multimodal learning methods typically treat different modalities equally by concatenation, which is equivalent to giving the fixed weight for each modality. In practical, item representation learning methods should be capable of assigning dynamic importance weights to each modality according to different items. Therefore, we propose a multimodal attention network to learn different weights of multiple modalities for each item, so that better item representations can be achieved by the weighted combination.

\subsection{Adversarial Transfer Learning}
Features of multiple modalities may contain redundant information, which should be eliminated when computing the different contributions of modalities. To exploit the redundant information, the common subspace across different modalities should be exploited. Adversarial transfer learning~\cite{ganin2015unsupervised,bousmalis2017unsupervised} is inspired by Generative Adversarial Nets~\cite{goodfellow2014generative}, which enables domain adaptation in deep networks that can be trained on a large amount of labeled data from the source domain and unlabeled data from the target domain.~\citet{wang2017adversarial} propose an adversarial cross-modal retrieval method, which seeks an effective common subspace based on adversarial learning.~\citet{pei2018multi} capture multimodal structures to enable fine-grained alignment of different data distributions based on multiple domain discriminators. Unlike previous adversarial transfer methods that solely match distribution at domain-level,~\citet{xie2018learning} propose to match distribution at class-level and align features semantically without any target labels.~\citet{yu2019transfer} propose a novel Dynamic Adversarial Adaptation Network to learn domain-invariant representations while quantitatively evaluate the relative importance of the marginal and conditional domain distributions.

The above adversarial learning methods are still a one-to-one adversarial paradigm, although the items in E-commerce involve multiple modalities. Moreover, different modality features involve different degrees of common features. Therefore, we propose a novel double-discriminators multimodal adversarial network to learn common latent subspace across multiple modalities.

\subsection{Context Aware Personalization Model}
In recent years, there have been growing numbers of researches on the personalization based on deep neural networks for recommending music~\cite{Oord2013}, news~\cite{oh2014personalized}, videos~\cite{covington2016deep}, and jobs~\cite{borisyuk2017lijar}.

Personalization for search and recommendation is typically based on user behaviors, where the RNN-based model is a good choice due to the sequential characteristic of user behaviors.~\citet{hidasi2015session} apply the RNN model to infer users' future intentions based on their previous click behavior sequence.~\citet{tan2016improved} present several extensions to the basic RNN model to enhance the performance of recurrent models.~\citet{ni2018perceive} adopt LSTM and the attention mechanism to model the user behavior sequence. Compared to sequence-independent approaches, these methods can significantly improve the CTR prediction accuracy and most of these techniques have been deployed in real-world applications~\cite{wu2016personal,ni2018perceive,zhou2018deep,zhou2019deep}. 

\begin{figure*}[t]
	\centering
	\vspace{-10mm}
	\includegraphics[width=1\linewidth]{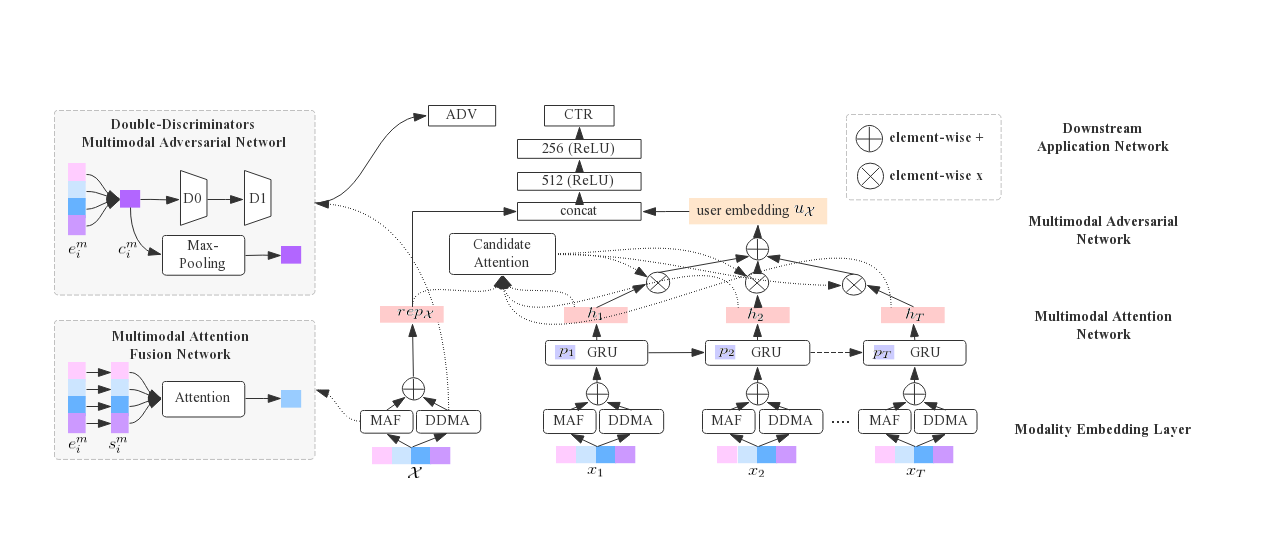}
	\vspace{-12mm}
	\caption{The architecture of MARN.}
	\label{fig:architecture}
	\vspace{-2mm}
\end{figure*}

\section{Our Proposed Method}
The typical framework of CTR prediction is to take a user behavior sequence $u=\{x_1,x_2,\ldots,x_T\}$ and the candidate item $\mathcal{X}$ as inputs, and aims to learn the probability that $u$ clicks $\mathcal{X}$. Items are mostly represented by ID to learn its representation through embedding. When considering multiple modalities, previous works either first extract multimodal embeddings and fuse these embeddings through concatenation~\cite{zhou2018deep,ni2018perceive,zhou2019deep}, or dynamically give different weights for each modality through an attention mechanism~\cite{li2018multi,wu2019neural}. Differently, we propose to divide multiple modality features into modality-specific and modality-invariant features. The overall architecture of the proposed MARN is presented in Figure~\ref{fig:architecture}. In our work, \textsl{Multimodal Attention Network} learns the dynamic weights of multiple modalities for each item according to its modality-specific features. \textsl{Multimodal Adversarial Network} learns common representations across multiple modalities for more comprehensive item representations. Besides the above two components, the \textsl{Modality Embedding Layer} and \textsl{Downstream Application Network} of our model are similar to related CTR prediction models. 

\subsection{Modality Embedding Layer}
Each item in the user behavior sequence is associated with a behavior property and forms a user-item interaction pair $<x_i,p_i>$. The modality embedding layer is applied upon the sparse features of each item in the user behavior sequence and its behavior property to compress them into low-dimensional vectors.

\subsubsection{Item Modality Embedding}
An item $x_i$ is represented by the multimodal information: i) \textbf{IDs}, unordered discrete feature including item ID, shop ID, brand ID and category ID; ii) \textbf{image}, pixel-level visual information; iii) \textbf{title}, word sequence; iv) \textbf{statistic}, historical exposure, click, transaction order/amount.

\textbf{IDs:} The IDs feature is represented as $[x_i^{id(1)},\ldots,x_i^{id(F)}]$, which is a multi-hot vector ($[\cdot, \cdot]$ denotes vector concatenation). $F$ is the number of IDs feature, and $x_i^{id(f)}$ is the $f^{th}$ feature which is a one-hot vector representing an ID like item ID or shop ID. The embedding layer transforms the multi-hot vector into a low-dimensional vector $e_i^{id}$ with an embedding lookup table, as shown in Equation~\ref{eq1}:

	\vspace{-5mm}
	\begin{equation}
	\label{eq1}
	\begin{split}
	e_i^{id} = &[W_{emb}^1x_i^{id(1)},\ldots, W_{emb}^Fx_i^{id(F)}], 
	W_{emb}^f\in R^{d_{emb}^f\times V_f}
	\end{split}
	\vspace{-7mm}
	\end{equation}
where $d_{emb}^f$ is the dimension and $V_f$ is the vocabulary size.

\textbf{Image:} Recent progress in computer vision shows that the learned semantic embeddings from the pixel-level visual information for classification tasks have good generalization ability~\cite{simonyan2014very,he2015delving}. Thus for an input image, the image embedding is the output of pre-trained VGG16~\cite{simonyan2014very} model with the last two layers for classification purpose removed, which result in a $4096$-dimensional vector.

\textbf{Title:} The title containing $h$ words (padded where necessary) is represented as an $h\times d_{emb}^{term}$ matrix, where each word is represented as a $d_{emb}^{term}$-dimensional vector. As suggested by Word2Vec~\cite{mikolov2013distributed}, we set $d_{emb}^{term}=300$. We design a fast convolutional network for title embeddings which uses multiple filters with varying window sizes $n=2,3,4$ to obtain the n-gram features, following ~\citet{kim2014convolutional}.

\textbf{Statistic:} We found it difficult in learning a good embedding directly on continuous statistic feature. Following~\citet{yuanfei2019autocross}, we adopt multi-granularity discretization, which discretizes each numerical feature into two, rather than only one, categorical features, each with a different granularity, ranging from 0 to 9 and 0 to 99 according to the number of items. We then perform categorial feature lookups to obtain two $8$-dimensional vectors.

\subsubsection{Behavior Property Embedding}
The behavior property $p_i$ describes the type and time of items in the user behavior sequence. Behavior type is a one-hot vector representing click, add-to-cart or purchase. Behavior time is a statistic feature indicating the seconds from the time it happens to the current recommendation time. We transform each behavior property into embedding $p_i=e_i^p$ and treat the the behavior property embedding $p_i$ as a strong signal to reflect the importance of each behavior.

\subsection{Multimodal Attention Network} \label{subsection_man}

Utilizing multiple modality features is often effective to improve the performance of CTR tasks. A straightforward way~\cite{zhou2018deep,ni2018perceive,zhou2019deep} is to concatenate the multiple modality features, which is equivalent to giving a fixed importance weight to each modality regardless of different items. A conceivable improvement~\cite{li2018multi,wu2019neural} is to dynamically distinguish the contributions of different modalities through an attention mechanism. However, features of multiple modalities may contain redundant information, which should be eliminated when computing the different contributions of modalities. 

Due to the above reason, we propose to explore the complementarity and redundancy of modalities by considering modality-specific and modality-invariant features differently. More precisely, the distributions of modality-specific features are separated, whereas the distributions of modality-invariant features are close to each other. To address this, we decompose each item embedding vector into modality-specific and modality-invariant features, i.e., $s_i^m$, $c_i^m$. More specifically, we project each item modality embedding into vector $e_i^m$ of the same dimension, and then apply two nonlinear projection layers to obtain two $256$-dimensional vectors according to $s_i^m,c_i^m=\mathcal{S}_m(e_i^m),\mathcal{I}(e_i^m)$. $\mathcal{S}_m(\cdot)$ is an independent projection matrix of each modality $m$ for extracting modality-specific features, while the projection matrix $\mathcal{I}(\cdot)$ is shared across modalities for learning modality-invariant features. 

\begin{figure}[t]
	\centering
	\vspace{-2mm}
	\includegraphics[width=1\linewidth]{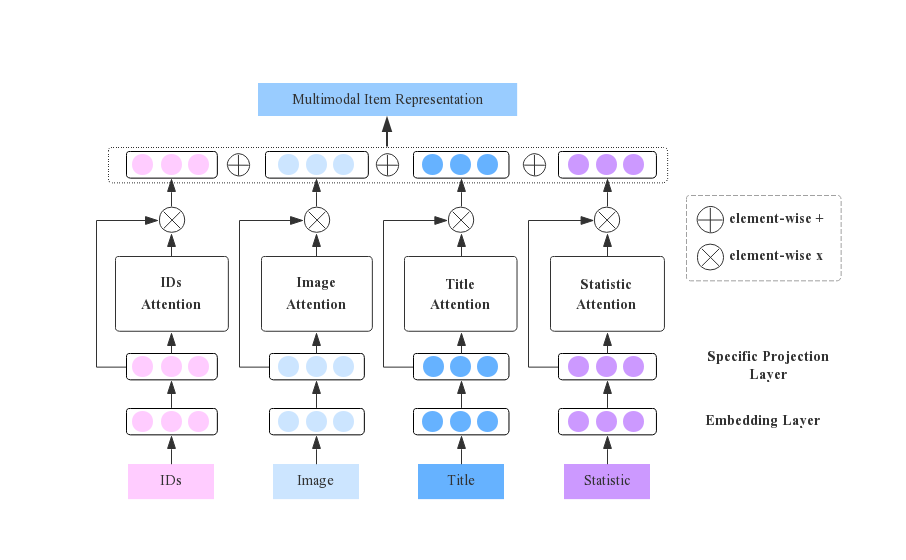}
	\vspace{-7mm}
	\caption{Illustration of multimodal attention fusion.}
	\label{fig:feature_level_fusion_net}
	\vspace{-0mm}
	\end{figure}

Therefore, we propose a multimodal attention fusion (MAF) network shown in Figure~\ref{fig:feature_level_fusion_net}, to learn dynamic weights according to its modality-specific features $s_{i}^m$ and then obtain the modality-specific representation $s_{i}$ by the weighted summation:

\vspace{-4mm}
\begin{eqnarray}
s_{i}&=&\sum_{m=1}^{M}atten_{i}^m\odot  s_{i}^m\\
atten_{i}^m&=&tanh(W_{m}^\top \cdot  s_{i}^m+b_{m})	
\vspace{-3mm}
\end{eqnarray}
where $atten_{i}^m$ controls the attention weight of modality $m$, and $\odot$ denotes element-wise multiplication. The parameters $W_{m}^\top$ is matrix with the size of $d\times d$, and  $b_{m}$ is a vector with the size of $d\times 1$, where $d=256$. The vector attention adjusts the importance weights of each dimension for the modality-specific features.

\subsection{Multimodal Adversarial Network}
Since the weights of the $\mathcal{S}_m(\cdot)$ and the $\mathcal{I}(\cdot)$ are only supervised by the label of the CTR task, we cannot expect that MARN has successfully learned the modality-specific and modality-invariant features. The distributions of modality-specific features extracted by $\mathcal{S}_m(\cdot)$ should be kept as far as possible, meanwhile, the distributions of modality-invariant features extracted by $\mathcal{I}(\cdot)$ should be drawn as close as possible. To address this, we design a multimodal adversarial network, which makes the modality-invariant feature extractor compete with a modality discriminator, in order to learn a common subspace across multiple modalities. Moreover, we propose a modality-specific discriminator to supervise the modality-specific feature extractor, so that the learned multimodal information has very good separability, leading to the modality-specific features.

\subsubsection{Cross-Modal Adversarial} 

The general cross-modal adversarial method~\cite{wang2017adversarial} seeks an effective common subspace across modalities, which is built around a minimax game, as:

\vspace{-2mm}
\begin{equation}
\begin{split}
     \underset{\mathcal{I}}{min}\underset{\mathcal{D}}{max}\mathcal{L}_{D}&=\mathbb{E}_{x\sim px^{1}}\left [log(\mathcal{D}(\mathcal{I}(x))) \right ]\\&+\mathbb{E}_{x\sim px^{2}}\left [log(1-\mathcal{D}(\mathcal{I}(x))) \right ]
\label{d_o}
\end{split}
\vspace{-4mm}
\end{equation}
where $px^{1}$ and $px^{2}$ are distributions of two modalities.

For $\mathcal{I}$ fixed, the optimal discriminator $D$ is:

\vspace{-2mm}
\begin{equation}
    D_\mathcal{I}^*(x)=\frac{px^{1}(x)}{px^{1}(x)+px^{2}(x)}
\label{d}
\end{equation}
\vspace{-1mm}

Given the optimum $D^*$, the minimax game of Equation~\ref{d_o} is:
\vspace{-0mm}
\begin{equation}
     \mathcal{L}_{D}=-log(4)+2\cdot JSD(px^{1}||px^{2})
\label{d}
\end{equation}
\vspace{-3mm}

Since the Jensen-Shannon divergence (JSD) between two distributions is always non-negative and achieves zero only when they are equal, $\mathcal{L}_{D}$ obtains the global minimum on $px^{1}=px^{2}$~\cite{goodfellow2014generative}, which is the goal of cross-modal adversarial learning.

\subsubsection{Double-Discriminators Multimodal Adversarial}

The above cross-modal adversarial method focuses on a one-to-one adversarial paradigm, while the items in E-commerce involve multiple modalities. Moreover, different modality features involve different degrees of common latent features, and the degrees can be treated as guidance to confuse the modality discriminator towards better discrimination performance. For example, the item ID embedding should be assigned relatively small weights due to the uniqueness. On the contrary, the image and title embeddings that involve much potential common subspace should be emphasized for further confusing the modality discriminator.

To address this, we propose a novel double-discriminators multimodal adversarial (DDMA) network. More precisely, we bring two contributions. First, we expand the original one-to-one adversarial paradigm to multimodal scenarios. Second, we introduce a double-discriminators strategy. The first discriminator identifies modality-invariant features that are potentially from the common latent subspace across multiple modalities and also emphasizes the identified modality-invariant features for further confusing the second discriminator. Meanwhile, the second discriminator drives knowledge transfer between modalities so as to learn common latent subspace across multiple modalities.

\begin{figure}[t]
	\centering
	\vspace{-2mm}
	\includegraphics[width=0.85\linewidth]{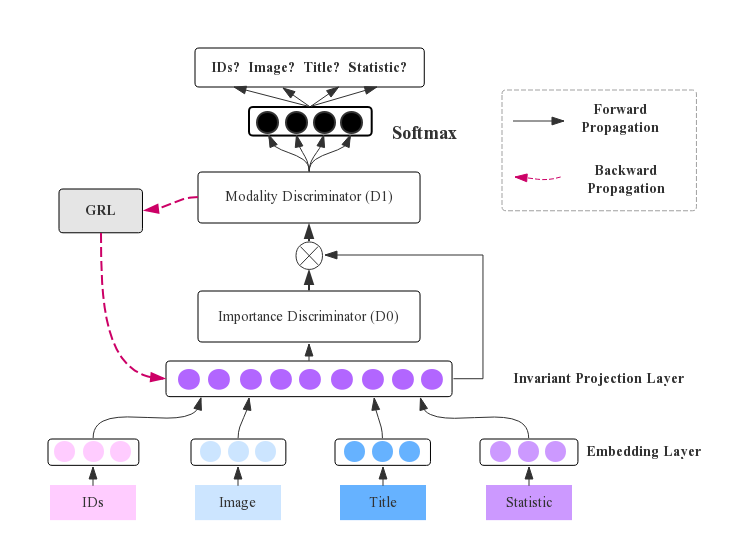}
	\vspace{-3mm}
	\caption{Illustration of double-discriminators multimodal adversarial network.}
	\label{fig:adv}
	\vspace{-5mm}
	\end{figure}

The first discriminator $D_0$ is an M-class classifier and is given by $D_{0}(x)=softmax(\mathcal{D}_0(x))$, where $x=c=\mathcal{I}(e)$, $e$ is the modality embedding, $\mathcal{I}$ is the invariant projection layer and $c$ is the projected modality-invariant feature. Suppose that $D_0$ has converged to its optimum $D_0^*$ and $x$ belongs to modality $i$, ${D}_0^i(x)$ gives the likelihood of $x$ in terms of modality $i$. If $D_{0}^{i*}(x)\approx 1$, then $x$ highly involves few common features across modalities, since it can be perfectly discriminated from other modalities. Thus, the degree of $x$ as $w^i(x)$ contributing to the modality-invariant feature should be inversely related to $D_{0}^i(x)$ according to $w^i(x)=1-D_{0}^i(x)$.

The second discriminator $D_{1}$ is also an M-class classifier, which is applied to reduce the JSD between modalities. After adding the degrees to the modality features for the discriminator $D_{1}$, the objective function of weighted multimodal adversarial network is:

\vspace{-3mm}
\begin{equation}
\begin{split}
    \underset{\mathcal{I}}{min}\underset{\mathcal{D}_1}{max}\mathcal{L}_{D_1}=\mathbb{E}_{e\sim pe^{m}}\sum_{m=1}^{M}\left [ w^m(x)log(\mathcal{D}_{1}(\mathcal{I}(e))) \right ] 
\label{min_max}
\vspace{-4mm}
\end{split}
\end{equation}
where $w^m(x)=w^m(\mathcal{I}(e))$ is a constant and independent of $\mathcal{D}_{1}$.

For fixed $\mathcal{I}$ and $\mathcal{D}_0$, the optimal output for $D_{1}^i$ is:

\vspace{-1mm}
\begin{equation}
    D_{1}^{i*}(x)=\frac{w^{i}(x)px^{i}(x)}{\sum_{m=1}^{M}w^{m}(x)px^{m}(x)}
\label{d_2}
\vspace{1mm}
\end{equation}

Given the optimum $D_1^*(x)$, the minimax game of Equation~\ref{min_max} is:

\vspace{-1mm}
\begin{equation}
\begin{split}
    \mathcal{L}_{D_1}=&\sum_{i=1}^{M}\!\!\int_x\!\! w^i(x)px^{i}(x)log(\frac{w^{i}(x)px^{i}(x)}{\sum_{m=1}^{M}w^{m}(x)px^{m}(x)})dx\\
=&\sum_{i=1}^{M}\!\!\int_x\!\! w^i(x)px^{i}(x)log(\frac{w_{i}(x)px^{i}(x)}{\sum_{m=1}^{M}\frac{w^{m}(x)px^{m}(x)}{M}})dx\!\!-\!\!Mlog(M)\\
=&\sum_{i=1}^{M}\!KL\!\left ( w^i(x)px^{i}(x)\parallel \sum_{m=1}^{M}\frac{w^{m}(x)px^{m}(x)}{M}\right )\!\!-\!\!Mlog(M)\\
=&\ M \!\cdot\! JSD_{\frac{1}{M},\cdots,\frac{1}{M} }\!\left (w^{1}(x)px^{1}(x),\! \cdots\!, w^{M}(x)px^{M}(x)\right )\!\!-\!\!Mlog(M)
\label{minmax_proof}
\end{split}
\vspace{-1mm}
\end{equation}

As deduced in Equation~\ref{minmax_proof}, the minimax game of Equation~\ref{min_max} is essentially reducing JSD between $M$ weighted modality features and obtains the optimum on $w^{1}(x)px^{1}(x)=\cdots=w^{M}(x)px^{M}(x)$.

Since $w^m(x)=w^m(\mathcal{I}(e^m))$ and $px^{m}(x)=c^m=\mathcal{I}(e^m)$, we can achieve modality-invariant representation through a max-pooling operation over the weighted modality-invariant features and take the maximum value, as $c_i=max\{w^1c^1,\ldots,w^Mc^M\}$.

\subsubsection{Modality-Specific Discriminator}

To encourage the modality-specific features $s_i^m$ to be discriminated between multiple modalities, we propose a modality-specific modality discriminator $D_s$ on the modality-specific features,where the discriminating matrix of ${D}_s$ is shared with the second discriminator ${D}_1$. 

Finally, we achieve the complementarity and redundancy of multiple modalities, resulting in the modality-specific representations $s_i$ and modality-invariant representations $c_i$. Then, the multimodal item representation $rep_i$ is produced by the element-wise summation of modality-specific and modality-invariant representations.

\subsection{Downstream Application Network}
\subsubsection{Behaviors Integrating Layer.}

With the proposed item representation learning model, we can integrate the user behaviors with various kinds of RNN-based model. In this paper, we choose GRU~\cite{hidasi2015session} to model the dependencies between the user behaviors. However, GRU lacks of capturing two characteristics: i) different behavior type and time contribute differently to extracting users' interests; ii) whether to click a candidate item or not highly depends on the most related behaviors with respect to the candidate item. Moreover, extracting the key information in the complex behaviors further contributes towards learning better item representations. 

Therefore, we propose an Attentional Property GRU (APGRU) layer. First, we modify the update gate of GRU to extract the most interested behavior state according to $u_t=\sigma (W^urep_t + P^up_t + U^uh_{t-1} + b^u)$, where $p_t$ is the behavior property, $rep_t$ is the $t^{th}$ item representation, and $\sigma$ is the sigmoid function.

Second, we concatenate the hidden state $h_t$ and the candidate item $rep_\mathcal{X}$ and apply a Multilayer Perceptron as the candidate attention network. Then we calculate the normalized attention weights, and the user embedding $u_\mathcal{X}$ can be calculated as the weighted summation of the output hidden states.

\subsubsection{CTR Prediction Layer}
In this paper, we set the prediction layer to be a deep network with point-wise loss for the CTR task. The overall objective functions are:
	\vspace{-0mm}
	\begin{eqnarray}
	\label{xx}
	\begin{split}
	&\underset{\mathcal{F},\mathcal{I},\mathcal{S}_m,\mathcal{E}}{min}\mathcal{L}_{CTR}= -\frac{1}{N}\sum_{n=1}^N [ y_n log (\mathcal{F}(u_n,rep_n))\\
	&\quad\quad\quad\quad\quad\quad\quad\quad+ (1-y_n)log(1-\mathcal{F}(u_n,rep_n)]\\
	&\underset{\mathcal{D}_s,\mathcal{S}_m}{min}\mathcal{L}_{D_s}=\lambda \ (-\mathbb{E}_{e,y\sim pe^{e,y}}\sum_{m=1}^{M}\mathbb{I}_{[ m=y]}log\mathcal{D}_s(\mathcal{S}_m(e)))\\
	&\underset{\mathcal{D}_0}{min}\mathcal{L}_{D_0}=-\mathbb{E}_{c,y\sim pc^{c,y}}\sum_{m=1}^{M}\mathbb{I}_{[ m=y]}log\mathcal{D}_0(c)\\
	&\underset{\mathcal{I}}{min}\underset{\mathcal{D}_1}{max}\mathcal{L}_{ADV}= \lambda \ (\mathbb{E}_{e\sim pe^{m}}\sum_{m=1}^{M}\left [ w^m(e)log(\mathcal{D}_{1}(\mathcal{I}(e))) \right ] )\\
	\end{split}
	\vspace{-0mm}
	\end{eqnarray}
where $u_n$ denotes the user embedding of the $n^{th}$ instance. $y_n$ denotes the label, which is set to 1 if user clicked item and 0 otherwise. $\lambda$ is a tradeoff hyper-parameter that controls the learning procedure of modality-specific and modality-invariant representations.

We optimize the four objective functions, i.e, $\mathcal{L}_{CTR}$, $\mathcal{L}_{D_s}$, $\mathcal{L}_{D_0}$, and $\mathcal{L}_{ADV}$ to update different layers in MARN simultaneously. First, we minimize $\mathcal{L}_{CTR}$ to optimize the overall prediction performance. Second, we minimize $\mathcal{L}_{D_s}$ to learn the modality-specific projection layers $\mathcal{S}_m$, while the gradient of $\mathcal{D}_s$ will not be back-propagated for updating the embedding layer $\mathcal{E}$. Then, we minimize the first discriminator $\mathcal{L}_{D_0}$ to identify the potential common subspace across multiple modalities. Since $\mathcal{D}_0$ are learned on un-weighted modality-invariant features and would not be a good indicator, the gradient of $\mathcal{D}_0$  will not be back-propagated for updating invariant projection layer $\mathcal{I}$. Finally, the second discriminator $\mathcal{D}_1$ plays the minimax game with the modality-invariant feature for updating $\mathcal{I}$. To solve the minimax game between $\mathcal{I}$ and $\mathcal{D}_1$, we adopt an end-to-end architecture through a gradient reversal layer (GRL)~\cite{ganin2015unsupervised}, so the sub-networks are trained jointly and boost each other.

\section{Experiments}
\begin{table*}[htbp]
    \vspace{0mm}
    \centering
    \begin{tabular}{ccccccccccccc}
    \toprule[0.75pt]
      Dataset & \# Users & \# Items & \# Shops & \# Brands & \# Categories & \# Images & \# Titles & \# Samples\\
    \hline
    Electro. (Amazon) & 192403 &63001 & - & -&801&63001 & 63001 & 2993570 \\
    Clothe. (Amazon)& 39387 &23033 & - & -&484&23033 & 23033 & 477862 \\
    Taobao & 0.25 billion & 0.8 billion & 17 million & 1.0 million & 20 thousand & 0.8 billion& 0.8 billion & 64 billion \\
    \bottomrule[0.75pt]
    \end{tabular}
    \caption{Statistics of Amazon and Taobao datasets.}
    \label{Taobao_stat}
    \vspace{-2mm}
\end{table*}

In this section, we present our experiments in detail, including dataset, evaluation metric, experimental setup, model comparison, and the corresponding analysis. Experiments on a public dataset with user behaviors and a dataset collected from Taobao Search system demonstrate MARN achieves remarkable improvements to the state-of-the-art methods on the CTR task. 

\subsection{Datasets and Experimental Setup}

\textbf{Amazon Dataset:} We use the Amazon dataset in~\citet{zhou2018deep}, which comprises several subsets of the amazon product data~\cite{mcauley2015image} and have sequential user behavior sequence. For a behavior sequence $\{x_1,x_2,\ldots,x_j,\ldots,x_T\}$, in the training phase, we predict the $j+1^{th}$ behavior with the first $j$ behaviors, where $j = 1,2,\ldots,T-2$. In the test phase, we use the first $T-1$ behaviors to predict the last one. The feature set we use contains item ID, category ID, image feature (extracted using pre-trained VGG16 model), statistic, together with $300$-dimensional GloVe vectors~\cite{pennington2014glove} for the terms of the title. The statistics of Amazon dataset is shown in Table~\ref{Taobao_stat}.

\textbf{Taobao Dataset:} We collect $8\times 10^9$ samples from the daily exposure and click logs in Taobao Search system, which consist of the user behaviors and the labels (i.e., exposure or click) for the CTR task. The user behavior type consists of click, add-to-cart and purchase. The feature set of an item contains item ID, shop ID, brand ID, category ID, image feature (extracted using pre-trained VGG16 model), statistic and title, also with a behavior type and time. For the production scenario, we use the samples across 7 days for training and evaluate on the samples of the next day. The statistics of Taobao dataset is shown in Table~\ref{Taobao_stat}.

We set the hyper-parameter configuration as follows. For Amazon and Taobao datasets, the dimensions of the item ID, shop ID, brand ID and category ID are set to be the same as in ~\cite{ni2018perceive}, which are 32, 24, 24 and 16, respectively. The vocabulary sizes of the embeddings are set according to the statistics of the datasets. Since we focus on multimodal item representation learning in this paper, we make the hidden units of the prediction layer as fixed values for all models, with [128,64] for the Amazon dataset and [512,256] for the Taobao dataset. The activation functions are ReLU. The model is trained on 32-sized batches for Amazon dataset and 1024-sized batches for Taobao dataset. We use AdaGrad~\cite{duchi2011adaptive} as the optimizer with the learning rate of 0.1. We gradually change $\lambda$ from 0 to $\lambda_0$ following the schedule $\lambda=\lambda_0(\frac{2}{1+\exp(-\gamma \cdot p)}-1)$, where $\gamma$ is set to 10 in all experiments (the schedule was not optimized/tweaked), and $p$ linearly changes from 0 to 1 in the training progress. The hyper-parameter $\lambda_0$ is tuned ranging from 0.01 to 1.8 and MARN achieves the best performance at $\lambda_0=0.05$.

\subsection{Offline Comparison of Different Methods}

To show the effectiveness of our method, we compare our method with three groups of nine baselines. The first group consists of models before deep networks for the CTR prediction task. 

\textbf{\textsl{LR}~\cite{mcmahan2013ad}:} Logistic regression (LR), which is a widely used shallow model before deep networks for CTR prediction task.

\textbf{\textsl{FM}~\cite{rendle2010factorization}:} Factorization Machine, which models both first-order feature importance and second-order feature interactions.

The second group contains concatenation-based mutlimodal fusion methods for mutlple modalities of items, which are the mainstream CTR prediction models.

\textbf{\textsl{YoutubeNet}~\cite{covington2016deep}:} A deep model proposed to recommend videos in YouTube, which gets user representations by simply averaging the item embeddings in the user behavior sequence.

\textbf{\textsl{xDeepFM}~\cite{lian2018xdeepfm}:} A compressed interaction network, which aims to automatically learn high-order feature interactions in both explicit and implicit fashions.

\textbf{\textsl{DUPN}~\cite{ni2018perceive}:} A general user representation learning method for E-commerce scenarios, which adopts LSTM and attention mechanism to model the user behavior sequence.

\textbf{\textsl{DIEN}~\cite{zhou2019deep}:} A two-layer RNN structure with attention mechanism. It uses the calculated attention values to control the second RNN layer to model drifting user interests.

The third group is formed of the state-of-the-art multimodal methods, which learn unified representations from multiple modalities and utilize variety of fusion techniques to improve the performance of the CTR prediction task. 

\textbf{\textsl{DMF}~\cite{hu2019dense}:} A hierarchically multimodal joint network, which densely correlates the representations of different modalities layer-by-layer, where the shared layer not only models the correlation in the current level but also depends on the lower one.

\textbf{\textsl{MMSS}~\cite{li2018multi}:} A modality-based attention mechanism model with image filters to selectively use visual information.

\textbf{\textsl{NAML}~\cite{wu2019neural}:} An attentive multi-view learning model, which incorporates titles, bodies and categories as different views of news, and applies attention mechanism to news encoder to select important views for learning informative news representations.

Note that, the inputs of LR and FM contains the IDs feature (item ID, shop ID, brand ID and category ID). All compared neural network models are fed with the same features as MARN for fair comparison. Finally, we conduct the significance test to verify the statistical significance of the performance improvement of our model against the baseline models.

\subsection{Evaluation Metrics}
Following~\cite{yan2014coupled,lian2018xdeepfm,ni2018perceive,zhou2019deep}, we adopt AUC as the evaluation metric, which is widely used in CTR prediction tasks. Average AUC is evaluated as following:

\vspace{-4mm}
\begin{equation}
  AUC=\frac{1}{\left | U^{Test} \right |}\sum_{u\in U^{Test}}\frac{1}{\left | I_u^+ \right |\left | I_u^- \right |}\sum_{i\in I_u^+}\sum_{j\in I_u^-}\delta (\hat{p_{u,i}}>\hat{p_{u,j}})
\vspace{-1mm}
\end{equation}
where $\hat{p_{u,i}}$ is the predicted probability that a user $u\in U^{Test}$ may click on the item $i$ in the test set and $\delta(\cdot )$ is the indicator function. $I_u^+$ is the item set clicked by the user, and $I_u^-$ is the item set that the user does not click. 

\begin{table}[t]
    \centering
    \begin{tabular}{lcccccccc}
    \toprule[0.75pt]
    \multirow{2}{*}{Group} &
    \multirow{2}{*}{Method} &
    \multicolumn{1}{c}{Electro.} &
    \multicolumn{1}{c}{Clothe.} &
    \multicolumn{1}{c}{Taobao} \\&
    & AUC  
    & AUC   
    & AUC \\  
    \hline
    \multirow{2}{*}{1}&\textsl{LR}  & 0.7272  & 0.7143  & 0.7011 \\
    &\textsl{FM}  & 0.7369  & 0.7224  & 0.7091 \\
    \hline
     \hline
    \multirow{6}{*}{2}&
    \textsl{YoutubeNet}  & 0.7675 & 0.7593& 0.7241 \\
    &\textsl{xDeepFM} & 0.7762 & 0.7691 & 0.7287\\
    &\textsl{DUPN} & 0.7904 & 0.7774 & 0.7348\\
    &\textsl{DIEN} & 0.7923 & 0.7793 & 0.7363\\
    \hline
     \hline
    \multirow{4}{*}{3}
    &\textsl{DMF} & 0.7924 & 0.7797 & 0.7371\\
    &\textsl{MMSS} & 0.7934 & 0.7806 & 0.7380\\
    &\textsl{NAML} & 0.7958 & 0.7825 & 0.7398\\
    &\textsl{MARN} & \textbf{0.8034*}&\textbf{0.7909*}&\textbf{0.7486*}\\
    \bottomrule[0.75pt]
    \end{tabular}
    \caption{AUC on Amazon and Taobao datasets. Here, * indicates statistical significance improvement compared to the best baseline (NAML) measured by t-test at $p$-value of 0.05. }
    \label{Amazon}
    \vspace{-7mm}
\end{table}

\subsection{Results on Amazon and Taobao Datasets}
All experiments are repeated 5 times and averaged metrics are reported in Table~\ref{Amazon}. The influence of random initialization on AUC is less than 0.0002. From Table~\ref{Amazon}, we can tell that MARN improves the performance significantly against all the baselines and achieves state-of-the-art performance. 

In the majority of the cases, non-neural network models in Group 1, i.e., LR and FM perform worse than neural network models, which demonstrates the power of deep learning. 

Comparing with the concatenation-based methods of Group 2, multimodal item learning methods of Group 3 outperforms all of them, which reflects the benefits of well designed multimodal fusion methods for learning good representation of items.

By comparing MARN with the models in Group 3, i.e., DMF, MMSS and NAML, although all the baselines are proposed to deal with multimodal information, MARN has better performance on the multimodal items in E-commerce. Since DMF was proposed to hierarchically joint multimodal information, it has the same issue as the concatenation-based methods which may ignore the different contributions of multiple modalites. MMSS was proposed to selectively use visual information with image filters, which may not utilize all image information effectively. Though the attention mechanism of NAML improves the performance, it has not explored the complementarity and redundancy of modalities by considering modality-specific and modality-invariant features differently. 

Finally, the improvement of MARN over the best baseline model NAML are 0.76\%, 0.84\% and 0.88\% AUC scores. As reported in~\cite{zhou2018deep}, compared with YoutubeNet, DIN improves AUC scores by 1.13\% and the improvement of online CTR is 10.0\%, which means a small improvement in offline AUC is likely to lead to a significant increase in online CTR. Considering the massive size of Taobao Search system, the lift of AUC scores by MARN may bring considerable CTR improvement and additional income each year.

\subsection{Ablation Study} \label{subsection_Application}
In this section, we design ablation experiments to study how each component in MARN contributes to the final performance. 

\textbf{\textsl{BaseModel}:} A model with unimodal item features which uses GRU to integrate the behavior embeddings. We conduct \textbf{ \textsl{BaseModel}+\textsl{IDs} } with IDs feature (item ID, shop ID, brand ID and category ID) and \textbf{\textsl{BaseModel}+\textsl{IMAGE}} with image feature.

\textbf{\textsl{BaseModel}+\textsl{CONC}:}  A concatenation model, where concatenation is performed in the fusion layer by element-wise summing up all the projected modality features (IDs, image, title and statistic).

\textbf{\textsl{BaseModel}+\textsl{MAF}:} A sub-model of MARN with the multimodal attention fusion network (MAF).

\textbf{\textsl{BaseModel}+\textsl{MAF}+\textsl{ADV}:} A sub-model of MARN with both MAF and the original adversarial transfer network (ADV)\cite{wang2017adversarial}.

\textbf{\textsl{BaseModel}+\textsl{MAF}+\textsl{DDMA}:} A sub-model of MARN with both MAF and the proposed double-discriminators multimodal adversarial network (DDMA).

\textbf{\textsl{MARN}:} The entire proposed model with MAF, DDMA and Attentional Property GRU (APGRU).

As shown in Table~\ref{application}, removing any component in MARN leads to a drop in performance. Further than that, we have the following observations. First, with the help of bringing in more modalities (i.e., image, title, and statistic), CONC outperforms the unimodal baselines by 0.35\%, 0.63\% and 0.65\% in AUC. Second, MAF outperforms CONC about 0.32\%, 0.35\% and 0.39\% in terms of AUC, which reflects that learning dynamic contributions of different modalities for multimodal item representations further leads to better performance. Third, both the original adversarial network and the proposed double-discriminators multimodal adversarial network can boost the CTR task performance, which reveals the benefits of identifying the potential common subspace across multiple modalities by multimodal adversarial learning. Moreover, the double-discriminators strategy (DDMA) outperforms ADV by 0.39\%, 0.51\% and 0.52\% AUC scores. It demonstrates that emphasizing the identified modality-invariant features by the first discriminator and further confusing the second discriminator are both crucial for learning better common latent subspace across multiple modalities. APGRU further provides AUC gain of 0.2\%, which results in the final performance.

\begin{table}[t]
    \centering
    \begin{tabular}{lccccccc}
    \toprule[0.75pt]
    \multirow{2}{*}{Method} &
    \multicolumn{1}{c}{Electro.} &
    \multicolumn{1}{c}{Clothe.} &
    \multicolumn{1}{c}{Taobao} \\
    & AUC  
    & AUC   
    & AUC  \\
    \hline
  \textsl{BaseModel}+\textsl{IDs}  & 0.7897& 0.7729& 0.7295 \\
    \textsl{BaseModel}+\textsl{IMAGE}  & 0.7631& 0.7564& 0.6833 \\
    \textsl{BaseModel}+\textsl{CONC} & 0.7929& 0.7792& 0.7360\\
    \hline
    \hline
    \textsl{BaseModel}+\textsl{MAF} &   0.7961&  0.7827&  0.7399 \\
    \textsl{BaseModel}+\textsl{MAF}+\textsl{ADV} &  0.7974&  0.7838&  0.7411  \\
    \textsl{BaseModel}+\textsl{MAF}+\textsl{DDMA} &  0.8013&  0.7889&  0.7463  \\
    \textsl{MARN} (\textsl{MAF}+\textsl{DDMA}+\textsl{APGRU}) & \textbf{0.8034}&\textbf{0.7909}& \textbf{0.7486}\\
    \bottomrule[0.75pt]
    \end{tabular}
    \caption{AUC on Amazon and Taobao datasets.}
    \label{application}
    \vspace{-7mm}
\end{table}

\begin{table*}[htbp]
    \centering
    \begin{tabular}{ccccccccccc}
    \toprule[1pt]
     Popularity-level   &0&1&2&3&4&5&6   &7&8&9\\
    \hline
    \textsl{BaseModel}+\textsl{IDs} & 0.672  & 0.687 & 0.694& 0.717& 0.725& 0.732& 0.736   &0.739& 0.750& 0.752\\
    \textsl{MARN} & 0.716  & 0.712 & 0.716& 0.738& 0.743& 0.745& 0.742   &0.741& 0.752& 0.753\\
    \textsl{GAP} & +4.4\%  & +2.5\% & +2.2\%& +2.1\%& +1.8\%& +1.3\%& +0.6\%  &+0.2\%& +0.2\%& +0.1\%\\
    \bottomrule[0.75pt]
    \end{tabular}
    \caption{AUC at different popularity levels on Taobao dataset.}
    \label{popularity}
     \vspace{-2mm}
\end{table*}

\begin{table*}[t]
    \centering
    \begin{tabular}{cccccccccccc}
    \toprule[1pt]
     Methods (top-$N$)    & 10& 20&30&40&50 & 60 &  70&  80&  90&  100&1000\\
    \hline
    \textsl{RANDOM}&0.01\%&0.01\%&0.02\%&0.02\%&0.03\%&0.04\%&0.05\%&0.06\%&0.07\%&0.07\%&0.59\%\\
    \textsl{BaseModel}+\textsl{IDs}&5.82\%&6.45\%&10.13\%&11.21\%&12.16\%&12.81\%&13.42\%&14.24\%&14.43\%&14.54\%&21.28\%\\
    \textsl{MARN}&6.95\%&7.97\%&12.31\%&13.63\%&14.86\%&15.34\%&15.59\%&15.75\%&15.86\%&15.99\%&21.99\%\\

    \bottomrule[0.75pt]
    \end{tabular}
    \caption{The click recall @top-N on Taobao dataset.}
    \label{unseen}
     \vspace{-2mm}
\end{table*}

\subsection{Item Representations Validation}

\subsubsection{Generalization on Unpopular Items.}
To verify what kinds of items benefit more from MARN, we uniformly divide all items in the test set of Taobao dataset into 10 equal frequency popularity levels according to their frequencies of interactions. The higher thepopularity level, the more popular items. We then compute the AUC at each level and show the results in Table~\ref{popularity}. It shows that both methods perform very well at high popularity levels from 7 to 9, while the performances decrease at low popularity levels from 0 to 3. For those less popular items, the unimodal model may not have enough training instances to capture the co-occurrence of the items. Fortunately, with the help of representing items with multimodal information, such as image and title features, MARN achieves extremely remarkable improvements at low popularity levels from 0 to 3. It demonstrates that \textsl{BaseModel}+\textsl{IDs} cannot deal with unpopular items well, while \textsl{MARN} boosts the generalization of the item representations and achieves better results.

\subsubsection{Transfer from Seen Items to Unseen Items.}
New items usually cause the cold-start problem, and many methods cannot deal with them. The item-based CF (Collaborative Filtering)~\cite{sarwar2001item} cannot perceive the items with no historical records, so the result are nearly random items (\textsl{RANDOM}). \textsl{BaseModel}+\textsl{IDs} can relieve the cold-start items using some generic IDs feature such as brand ID and shop ID. MARN can further address the cold-start problem by constructing an item embedding vector from not only generic IDs feature, but also image and title features. To study item transferability on unseen items, we evaluate the click recall@top-$N$ of the generated candidate set in the next day. The results are shown in Table~\ref{unseen}. The classical item-based CF cannot deal with new items. Compared to \textsl{BaseModel}+\textsl{IDs}, MARN achieves the improvement of about 2\% when $N$ is less than 70 and about 1\% when $N$ is beyond 70. Although the effectiveness decreases as $N$ increases, this is rare case because the number of items browsed by users is usually limited.

\subsection{Case Study of Multimodal Attention} 

\begin{figure}[t]
	\centering
	\vspace{-2mm}
	\includegraphics[width=1\linewidth,height=3.2cm]{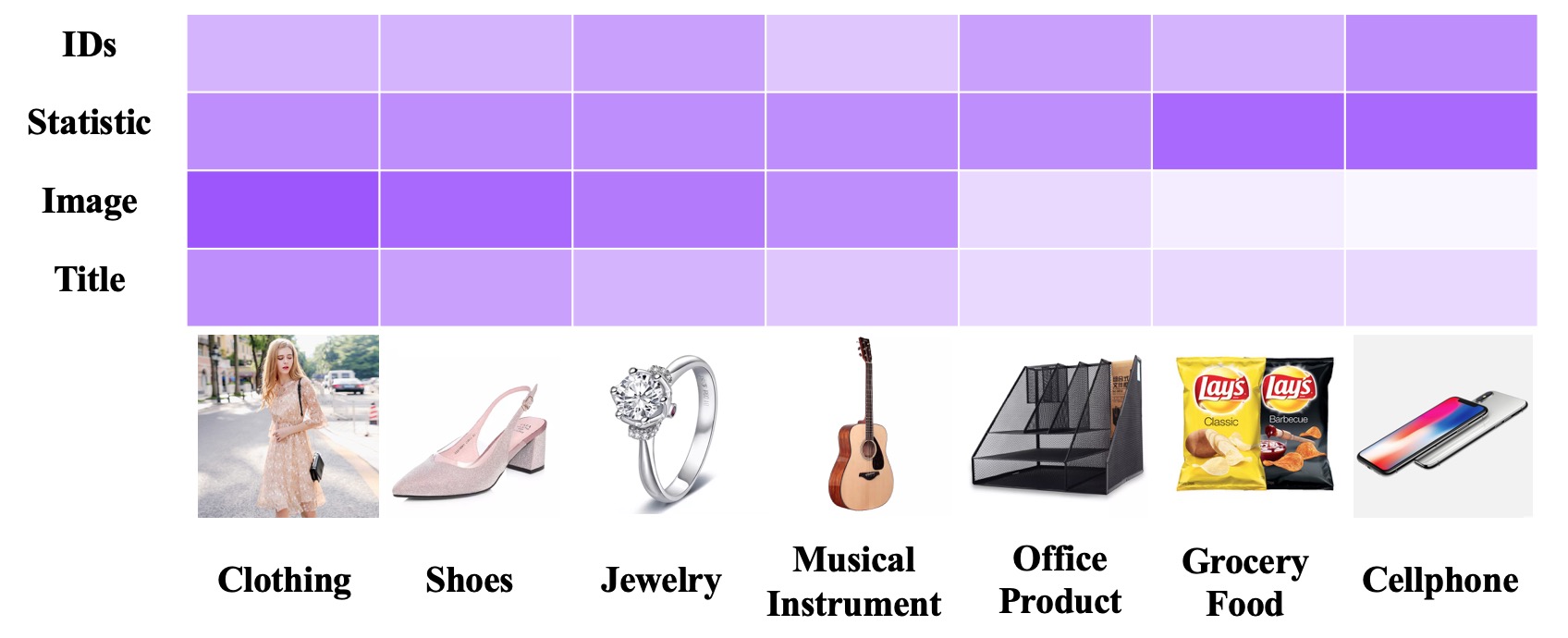}
	\vspace{-5mm}
	\caption{The heat map of attention weights with different item modalities of different categories.}
	\label{fig:attention}
	\vspace{-2mm}
\end{figure}

To verify the effectiveness of the multimodal attention in our model, we visualize several categories in Taobao dataset in Figure~\ref{fig:attention}. We randomly select ten thousand items for each category and each item is represented by multiple modalities including IDs, image, title and statistic. We then calculate the average $l2$-norm of the vector attention of multiple modalities for each category. The heat map matrix represents the attention intensity of each modality with respect to different categories. 

We can find that the image feature is highly weighted under categories such as clothing, shoes, and jewelry, while the statistic feature gains more weight in the cell phone and grocery food categories than that of the image feature. As a result, the multimodal attention mechanism in MARN can effectively learn dynamic contributions of different modalities for multimodal item representations.

\subsection{Empirical Analysis of Adversarial Network}

\subsubsection{Modality Visualization.} 
\begin{figure}[t]
\centering
\vspace{-0mm}
\includegraphics[width=0.9\linewidth]{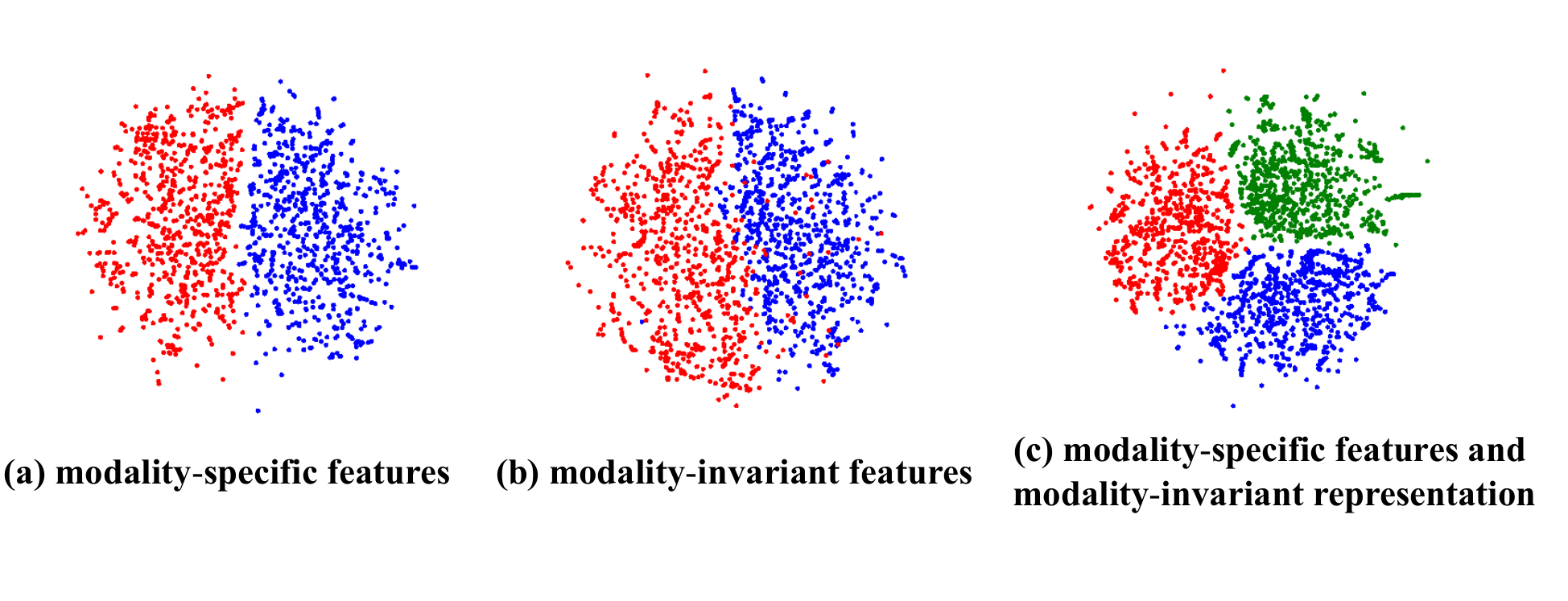}
\vspace{-6mm}
\caption{The t-SNE modality visualization. Red points denote the image modality feature,  while blue points denote title modality feature. Green points denote the modality-invariant representation obtained by max-pooling.}
\label{tsne}
\vspace{-1mm}
\end{figure}

To verify the effectiveness of the proposed multimodal adversarial network, we visualize the t-SNE features for ten thousand randomly selected items in Taobao dataset. The visualized features are modality-specific features ($s^{image}_{i}$ and $s^{title}_{i}$), modality-invariant features ($c^{image}_{i}$ and $c^{title}_{i}$) and modality-invariant representation ($c_{i}$). As shown in Figure\ref{tsne}(a), $s^{image}_{i}$ and $s^{title}_{i}$ are almost completely separated, which reflects that the auxiliary modality discriminator can reduce the redundancy and increase the distinctiveness of the modality-specific features. As shown in Figure\ref{tsne}(b), $c^{image}_{i}$ and $c^{title}_{i}$ are drawn close to each other, which reveals that the multimodal adversarial mechanism has the ability to decrease the modality gap and align distributions of different modalities. Figure\ref{tsne}(c) illustrates that $s^{image}_{i}$, $s^{title}_{i}$ and $c_{i}$ are completely separated, which means MARN can effectively exploit the complementarity and redundancy of multiple modalities and achieve universal item representations by combining both modality-specific and modality-invariant representations.

\begin{figure}[t]
\centering
\vspace{-0mm}
\subfigure[Convergence.]{
\label{loss}
\begin{minipage}[t]{0.5\linewidth}
\centering
\includegraphics[width=0.85\linewidth]{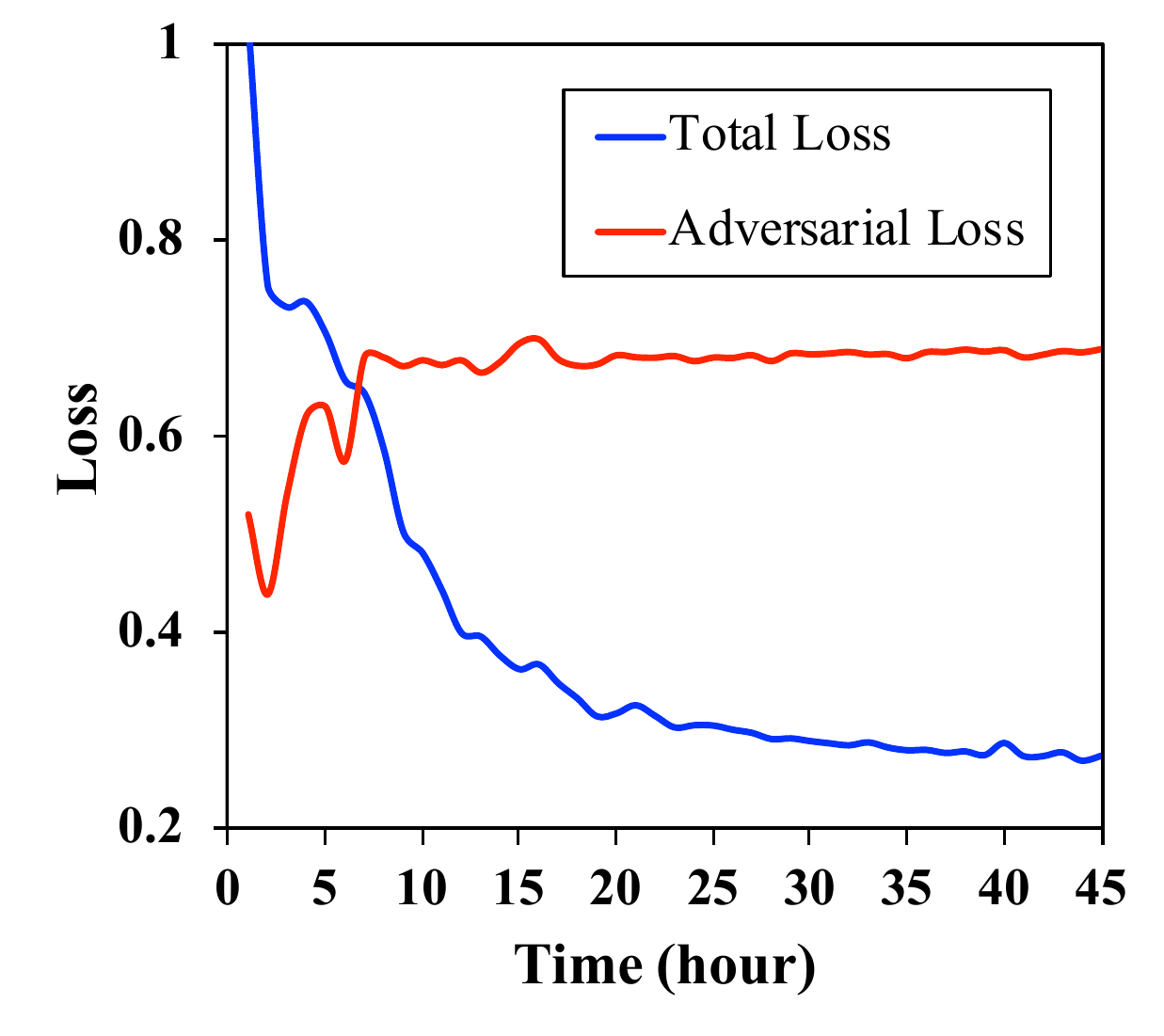}
\end{minipage}
}%
\subfigure[Parameter Sensitivity.]{
\label{param}
\begin{minipage}[t]{0.5\linewidth}
\centering
\includegraphics[width=0.85\linewidth]{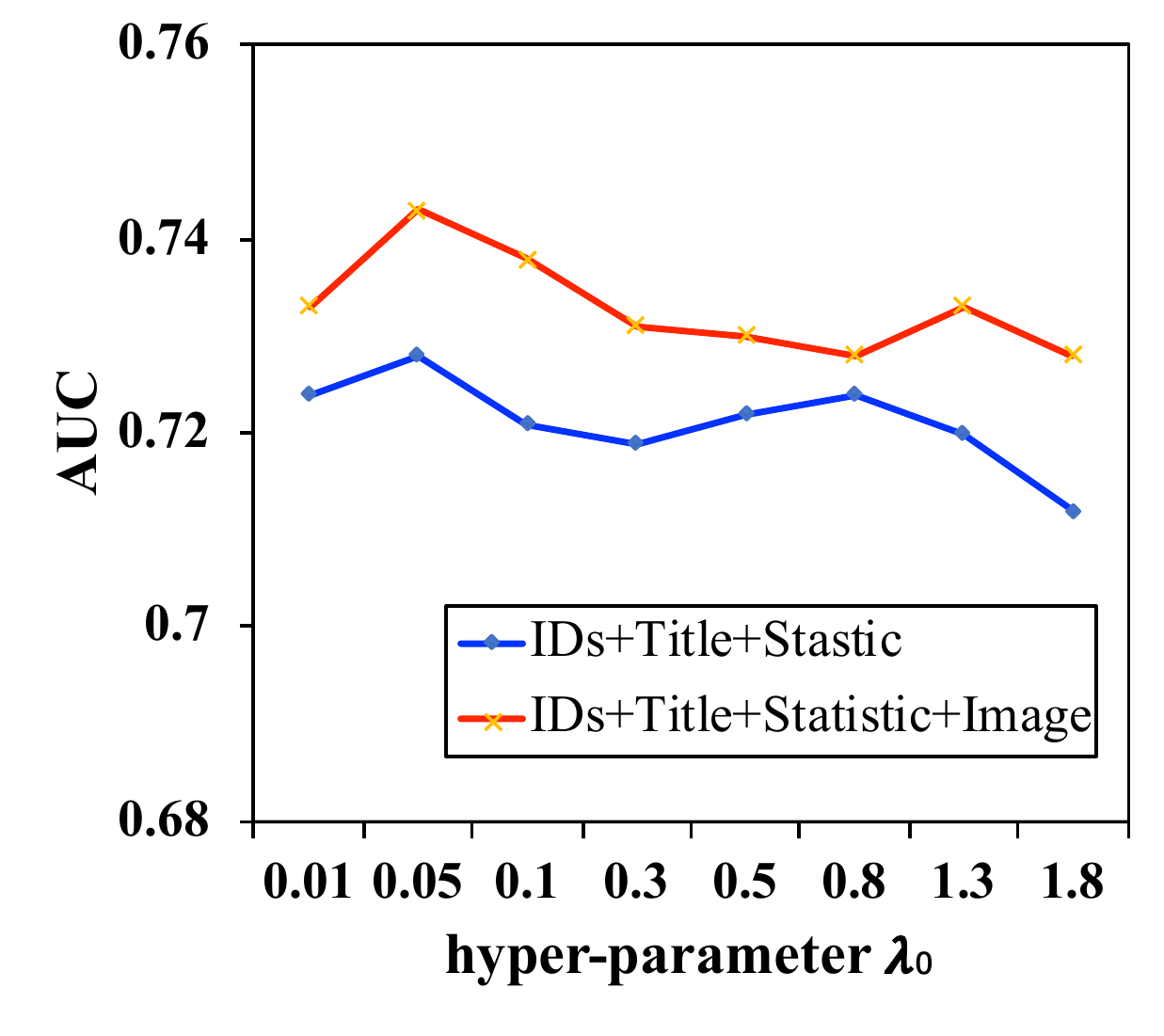}
\end{minipage}%
}
\centering
\vspace{-4mm}
\caption{Convergence and parameter sensitivity study.}
\vspace{-0mm}
\end{figure}

\subsubsection{Convergence and Parameter Sensitivity.} As shown in Figure~\ref{loss}, along with the training procedure, the total loss of MARN on Taobao dataset converges to a stable state, while the adversarial loss gradually increases, which means the modality discriminator becomes incapable of discriminating between different modalities based on the learned modality-invariant features. Figure~\ref{param} shows the AUC of MARN on Taobao dataset with the hyper-parameter $\lambda_0$ ranging from 0.01 to 1.8. MARN achieves the best performance at $\lambda_0=0.05$, which is an appropriate balance.

\subsection{Online Evaluation}
\subsubsection{Online Metrics.}
Careful online A/B testing in Taobao Search system was conducted for one month. The evaluation metrics are computed over all exposure items returned by the ranking system based on the CTR prediction models (i.e., YoutubeNet, DUPN and MARN). We report two metrics: CTR and GMV (Gross Merchandise Volume). CTR is computed as the ratio between the number of clicked items and the total number of exposure items. GMV is computed by the transaction number of items multiplied by the price of each item.

\begin{table}[h]
\vspace{-2mm}
\centering
\begin{center}
\begin{tabular}{ ccc}
\toprule[0.75pt]
Method & CTR Improve & GMV Improve\\
\hline
\textsl{YoutubeNet} & 0\% & 0\% \\
\textsl{DUPN} & 5.23\% & 3.17\% \\
\textsl{MARN} & \textbf{10.46\%} & \textbf{5.43\%}\\
\bottomrule[0.75pt]
\end{tabular}
\end{center}
\caption{Online A/B testing.}
\label{Online A/B Test}
\vspace{-7mm}
\end{table}

Due to the policy restrictions, we do not report the absolute numbers for the online metrics. Instead, we report the relative numbers with respect to \textsl{YoutubeNet}. As shown in Table \ref{Online A/B Test}, compared to \textsl{DUPN}, the last version of our online serving model, \textsl{MARN} has improved CTR by 5.23\% and GMV by 2.26\%. Considering the massive size of Taobao Search system, such consistent online improvements are significant. MARN has already been deployed online, serving the main traffic for hundreds of million users with billions of items.

\subsubsection{Online Serving.}
Online serving of industrial deep networks is not an easy job with hundreds of millions of users visiting our system everyday. Even worse, at traffic peak our system serves more than 200,000 users per second. It is required to make real-time CTR predictions with high throughput and low latency. Though MARN is complex, it has additional advantage that the multimodal item representations can be pre-extracted during the stage of model deploying. Since MARN automatically learns the weights of different modalities only according to the item itself, we can perform the sub-network of MARN to extract the multimodal representation of each item and store the representation vector in the index of the search engine, thereby reducing the latency of online serving. Moreover, several important techniques are deployed for accelerating online serving of industrial deep networks: i) quantization for recurrent neural networks, which adopts multi-bit quantization strategies to accelerate online inference; ii) heterogeneous calculations including GPUs and ALI-FPGAs, which can accelerate matrix computations. In conclusion, optimization of these techniques doubles the QPS (Query Per Second) capacity of a single machine practically. Online serving of MARN also benefits from this.

\section{Conclusions}
This paper proposes a novel multimodal representation learning method for multimodal items, which can improve CTR and further boost GMV in E-commerce. We explore the complementarity and redundancy of multiple modalities by considering modality-specific and modality-invariant features differently. To achieve discriminative representations, we propose a multimodal attention fusion network. Moreover, to achieve a common latent subspace across modalities, we propose a double-discriminators multimodal adversarial network. We perform extensive offline experiments on Amazon and Taobao datasets to verify the effectiveness of the proposed method. MARN consistently achieves remarkable improvements to the state-of-the-art methods. Moreover, the approach has been deployed in an operational E-commerce system and online A/B testing further demonstrates the effectiveness.

\section{Acknowledgement}
We thank colleagues of our team - Zhirong Wang, Tao Zhuang, Heng Li, Ling Yu, Lei Yang and Bo Wang for valuable discussions and suggestions on this work. We thank our search engineering team for the large scale distributed machine learning platform of both training and serving. We also thank scholars of prior works on multimodal representation learning and recommender system. We finally thank the anonymous reviewers for their valuable comments.

\clearpage
\balance
\bibliographystyle{ACM-Reference-Format}
\bibliography{sample-bibliography}
\end{document}